\documentclass[prl,aps,tightenlines,twocolumn]{revtex4}
\usepackage[]{graphicx}
\usepackage[]{epsfig}
\begin{document}

\title{Self-referenced opto-mechanical oscillator}

\author{A. B. Matsko, A. A. Savchenkov, V. S. Ilchenko, D. Seidel, and L. Maleki}

\affiliation{OEwaves Inc., 2555 East Colorado Blvd. Ste. 400, Pasadena, CA 91107}

\begin{abstract}
We propose a method for frequency stabilization of a resonant opto-mechanical oscillator utilizing the different temperature and strain dependence of the optical and mechanical modes. In particular, we show that the temperature of the mechanical resonator can be stabilized to  microKelvin level without the need for an external temperature or frequency reference. The method is promising for reaching the thermodynamically limited performance of opto-mechanical oscillators, and even for lifting the thermodynamic limit.
\end{abstract}

\maketitle

Recent developments in the field of cavity opto-mechanics brought to life theoretical ideas from decades ago that suggested macroscopic mechanical devices could be prepared in quantum-limited ground states \cite{kippenberg08s}. In-depth studies of opto-mechanical systems also resulted in creation of novel high-stability regenerative photonic oscillators based on  opto-mechanical parametric instability \cite{hosseinzadeh06pra,rokhsari06apl,hosseinzadeh08apl}. These oscillators are interesting from both fundamental and practical points of view, as they allow investigating opto-mechanical interactions with precision previously unachievable, as well as allowing generation of spectrally pure radio-frequency (RF) signals.

Opto-mechanical oscillators (OMO) are usually based on interaction of an optical and a mechanical mode in an optical resonator with mechanical degrees of freedom. The value of the OMO linewidth is ultimately limited by the Schawlow-Townes-like formula
\begin{equation} \label{ST}
\delta \omega_{c\ ST} \geq \frac{\gamma_M}{{\bar n}_c} ({\bar n}_{th}+1),
\end{equation}
where $\delta \omega_{c\ ST}$ is the linewidth of the oscillator, $\gamma_M$ is  half width at half maximum (HWHM) of the cold (un-pumped) mechanical mode, ${\bar n}_c$ is the averaged number of phonons in the mechanical mode, $\bar {n}_{th} = [\exp(\hbar \omega_c/k_BT) -1]^{-1}$ is the averaged number of phonons from the thermal bath in the mechanical mode. The linewidth decreases inversely with the number of phonons ${\bar n}_c$ generated in the mechanical mode, as shown in \cite{vahala08pra}. It is possible to retrieve the information about  mechanical oscillations using the light leaving OMO and, hence, use the OMO in practice.

One of disadvantages of existing OMOs is that the frequency stability of the mechanical mode is determined  in the same way as the stability of the electronic, e.g. quartz, oscillators. The fluctuations of the thermal bath as well as the thermodynamical noise and drifts of the mechanical mode limit the frequency stability. The pumping light is used as the power source in an OMO, similar to the electric power source in electronic oscillators and, therefore, it is unlikely that OMO will outperform its electronic analog. The optical nature of the oscillator is not utilized to its full capacity.

The goal of the present contribution is to show how the significant difference between  optical and mechanical frequencies can be used to stabilize the oscillation frequency of an OMO. We consider a triply resonant OMO based on interaction of one mechanical and two optical modes. The modes interact if the mechanical frequency corresponds to the frequency difference between two high-Q optical modes, and if the overlap integral of the modes is nonzero. Resonant scattering of light on surface acoustic waves is an example of such an interaction \cite{matsko09prl}. The oscillation occurs if the higher frequency optical mode is pumped with coherent light and the power of the pump exceeds a certain threshold. As a result of oscillation, the pump photon is divided into a confined mechanical phonon and a photon generated in the other optical mode that is red shifted with respect to the pump,.

The stabilization of the OMO is possible since there exists an opto-mechanical interaction that is phase matched in a very narrow spectral region and the power of the red-shifted light strongly depends on the phase matching. Phase matching changes as the temperature of the opto-mechanical resonator drifts. The change is negligible if the oscillation frequency is small and the pump and scattering light coexist in one optical mode \cite{hosseinzadeh06pra,rokhsari06apl,hosseinzadeh08apl}. However, the thermal dependence becomes very strong if the mechanical frequency significantly exceeds the linewidth of  optical modes, and two optical modes belonging to different mode families are involved in the process. We show, that in some resonators  phase matching occurs in a temperature range smaller than a milliKelvin. It is possible to keep the temperature of the resonator at the microKelvin level, and feeding back the information about the power of the generated signal to a thermal controller for the resonator. This temperature locking mechanism does not require an external temperature or a frequency reference. The technique also results in stabilization of the oscillation frequency with respect to stress and strain of the resonator, since modes involved in the oscillation process respond differently to mechanical deformation of the system.

In what follows we describe properties of the triply-resonant OMO and suggest a way of thermal stabilization of the oscillator using a particular example of a crystalline whispering-gallery mode (WGM) opto-mechanical resonator. We consider a WGM resonator made out of z-cut lithium niobate and study the case of coupling  TE (extraordinary) and TM (ordinary) optical WGMs, together with a surface acoustic wave (SAW) mode \cite{matsko09prl}. We show that  temperature of the mechanical mode can be stabilized to a microKelvin level in this particular system.

The triply-resonant opto-mechanical interaction is described by equations
\begin{eqnarray} \label{q1}
&& \dot  A = -\Gamma_A A - ig    C   B +  F_A,
\\
\label{q2} && \dot {  B} = - \Gamma_B   B -i g   C^*    A,
\\
\label{q3} && \dot{  C} = -  \Gamma_C   C -  i g    B^*
A .
\end{eqnarray}
where $  A$, $  B$, and $  C$ are the slowly-varying
amplitudes pump (optical), idler (optical mode red shifted with respect to the pump), and signal (mechanical) fields; $\Gamma_A$, $\Gamma_B$, and  $\Gamma_C$ are the linear resonant terms of optical and mechanical modes respectively
\begin{eqnarray} \nonumber
&& \Gamma_{A} =  i (\omega_a -\omega_0) + \gamma + \gamma_{ca},
\\
\nonumber && \Gamma_{B} =  i (\omega_b - \omega_-) + \gamma + \gamma_{cb},
\\
\nonumber && \Gamma_C = i (\omega_c - \omega_M) + \gamma_M,
\end{eqnarray}
$\gamma$ and $\gamma_M$ are the intrinsic decay rates of the optical and mechanical modes,
$\gamma_{ca}$ and $\gamma_{cb}$ are optical loading (coupling) rates (he loading of the optical modes can be different because modes belong to different families); $g$ is the acousto-optical coupling constant,
\begin{equation}
g=\omega_0\sqrt{\frac{K_\epsilon \hbar }{2m^*R^2 \omega_c}},
\end{equation}
$K_\epsilon$ is the correction coefficient showing that radiation pressure results in a changein the size of the resonator, and in its index of refraction, through strain.
$  F_A$ stands for  optical pumping
\begin{eqnarray}
  F_A = e^{i\phi_A}\sqrt{\frac{2P\gamma_{ca}}{\hbar \omega_0 }},
\end{eqnarray}
$P$ is the power of the external optical pump of  mode $A$.

We assume that
\begin{eqnarray}
  A  = |   A  | e^{i \phi_A}, \\
  B  = |   B  | e^{i \phi_B}, \\
   C  = |   C  | e^{i \phi_C},
\end{eqnarray}
and derive from Eqs.~(\ref{q2}-\ref{q3}) in the steady state
\begin{eqnarray}&&
\frac{|   B  |^2}{|   C  |^2} = \frac{\gamma_M}{\gamma + \gamma_{cb}}, \label{s1} \\ && \frac{\omega_b - \omega_- }{\omega_c - \omega_M } = \frac{\gamma + \gamma_{cb}}{\gamma_M}, \\
&& e^{i(\phi_A-\phi_B-\phi_C-\pi/2)}=e^{i \phi_{\Gamma_B}}, \\ &&
|   A  |^2= \frac{\gamma_M}{\gamma + \gamma_{cb}} \frac{|\Gamma_B|^2}{g^2}.
\label{s4}
\end{eqnarray}
In accordance with Eq.~(\ref{s4}) the pump field is clumped inside the WGM resonator as soon as the threshold is reached. Below threshold, $   A =   F_A /\Gamma_A$, the above threshold Eq.~(\ref{s4}) is valid. It is useful to introduce a dimensionless parameter $\xi$ and write the threshold condition in the form
\begin{equation} \label{xi}
\xi=\frac{|\Gamma_A||\Gamma_B|}{g|   F_A  |} \sqrt{\frac{\gamma_M}{\gamma + \gamma_{cb}}}<1.
\end{equation}
The condition basically means that the pump amplitude is always less than the pump amplitude with no oscillation occuring.

At the next step, we find from Eq.~(\ref{q1})
\begin{eqnarray} \label{s5}
&& g |   B  |^2 \sqrt{\frac{\gamma + \gamma_{cb}}{\gamma_M}} \cos(\phi_{\Gamma_A}+\phi_{\Gamma_B})= \\ \nonumber && |   F_A | \cos(\phi_{F_A}-\phi_{A}-\phi_{\Gamma_A})-\frac{|\Gamma_A||\Gamma_B|}{g} \sqrt{\frac{\gamma_M}{\gamma + \gamma_{cb}}}, \\ && g | B  |^2 \sqrt{\frac{\gamma + \gamma_{cb}}{\gamma_M}} \sin(\phi_{\Gamma_A}+\phi_{\Gamma_B})+\nonumber \\  && |   F_A  | \sin(\phi_{F_A}-\phi_{A}-\phi_{\Gamma_A})=0; \label{s6}
\end{eqnarray}

To analyze the behavior of the system we rewrite Eqs.~(\ref{s5}) and (\ref{s6}) as
\begin{eqnarray} \label{s5a}
&& \cos(\phi_{F_A}-\phi_{A}-\phi_{\Gamma_A})+ \\ \nonumber && {\rm ctg} (\phi_{\Gamma_A}+\phi_{\Gamma_B})\sin(\phi_{F_A}-\phi_{A}-\phi_{\Gamma_A})= \xi, \\
&& \zeta=\frac{g | B  |^2}{|   F_A  |} \sqrt{\frac{\gamma + \gamma_{cb}}{\gamma_M}} = -\frac{\sin(\phi_{F_A}-\phi_{A}-\phi_{\Gamma_A})}{\sin(\phi_{\Gamma_A}+\phi_{\Gamma_B})}. \label{s6a}
\end{eqnarray}
Eq.~(\ref{s5a}) can be solved analytically using substitution
\begin{equation}
\sin(\alpha)= \frac{2 {\rm tg}(\alpha/2) }{1+{\rm tg}^2(\alpha/2)}, \;\;
\cos(\alpha)= \frac{1-{\rm tg}^2(\alpha/2) }{1+{\rm tg}^2(\alpha/2)},
\end{equation}
where $\alpha=\phi_{F_A}-\phi_{A}-\phi_{\Gamma_A}$;
and
\begin{eqnarray}
{\rm tg}(\alpha/2)=\frac{{\rm ctg} (\phi_{\Gamma_A}+\phi_{\Gamma_B})}{1+\xi} \\ \nonumber
\pm \sqrt{ \left [ \frac{{\rm ctg} (\phi_{\Gamma_A}+\phi_{\Gamma_B})}{1+\xi}  \right ]^2 +\frac{1-\xi}{1+\xi}            }.
\end{eqnarray}

Using expressions
\begin{equation}
\omega_{0}=\omega_{-}+\omega_{M}, \;\; \omega_a-\omega_0=\Delta_a,
\end{equation}
we parameterize the resonance coefficients as
\begin{eqnarray} \nonumber
&& \Gamma_{A} =  i \Delta_a + \gamma + \gamma_{ca},
\\
\nonumber && \Gamma_{B} =  i \frac{\displaystyle \frac{\gamma + \gamma_{cb}}{\gamma_M} (\Delta_c+\Delta_a)}{1+\displaystyle \frac{\gamma + \gamma_{cb}}{\gamma_M}} + \gamma + \gamma_{cb},
\end{eqnarray}
where $\Delta_a$ is a free parameter given by the laser tuning with respect to the pump mode, and $\Delta_c=\omega_c+\omega_b-\omega_a$ is determined by relative change of optical and mechanical frequencies. We assume that the laser is locked to the pumping mode with certain fixed detuning $\Delta_a$. This detuning hardly changes when the temperature of the resonator changes as the laser follows the mode; however, the opto-mechanical detuning $\Delta_c$ changes rather significantly. The change of $\Delta_c$ results in a phase shift of the pumping light exiting the resonator with respect to the pumping light entering the resonator, as well as in a change of the amplitude of generated sideband.

The frequency instability of the oscillator results primarily from  technical temperature fluctuations of the resonator. The temperature must be stabilized to achieve an acceptable level of oscillation stability. We propose to stabilize the oscillator using the temperature dependence of  phase matching in the opto-mechanical process.  The spectral width of the gain of the process is given by the bandwidth of  optical modes and the power of the optical pump. If the mode bandwidth is narrow, the gain bandwidth is narrow as well. This allows stabilizing the temperature of the resonator as the small temperature drift moves the process out of phase matching and the system stops oscillating.

\begin{figure}
\centering\includegraphics[width=5cm]{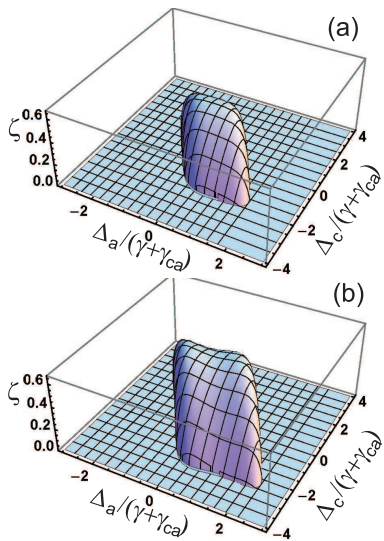}
\caption{\label{fig1} Normalized amplitude of the generated optical sideband (parameter $\zeta$, Eq.~\ref{s6a}) for moderate pump power $\xi=0.4$ (a) and $\xi=0.5$ (b) (the oscillation threshold corresponds to $\xi=1$ and  decreases with an optical power increase). The mechanical resonance is assumed to have a smaller bandwidth compared with the optical resonance $\gamma_M/(\gamma + \gamma_{ca})=0.1$. }
\end{figure}

\begin{figure}
\centering\includegraphics[width=8cm]{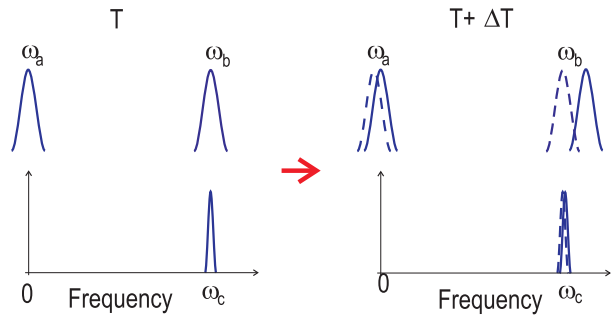}
\caption{\label{fig2} Illustration of the impact of  thermal shift on the mode structure of the resonator. We assume that at some initial temperature $T$ the difference in  optical frequencies corresponds to the frequency of the mechanical mode (left hand side in frequencies of  optical modes such that their frequency difference becomes substantially different from the frequency of the mechanical mode (right hand side). The mechanical mode experiences the thermal shift as well, but this is generally of much smaller magnitude  to influence the differential detuning. Dashed line figures in the right hand side of the picture show the initial position of  modes, while solid line figures stand for the final position.}
\end{figure}

Let us assume that  optical modes have identical loaded Q-factors $\gamma + \gamma_{cb}=\gamma + \gamma_{ca}$, and plot the normalized photon number in the optical sideband $\zeta$ (\ref{s6a}) as a function of $\Delta_a$ and $\Delta_c$ (see Fig.~\ref{fig1}).
A change in the temperature by $\Delta T$ results in a change of  optical and mechanical frequencies (Fig.~\ref{fig2})
\begin{eqnarray}
\Delta \omega_a = \beta_e \Delta T \omega_a, \\
\Delta \omega_b = \beta_o \Delta T \omega_b, \\
\Delta \omega_c = \beta_{s} \Delta T \omega_c.
\end{eqnarray}
where $\beta_e$, $\beta_o$, and $\beta_{s}$ describe the integral temperature dependence of the mode frequency (we assume that the resonator is made out of z-cut uniaxial material). For optical modes the coefficient includes a shift in the mode frequency due to thermal expansion and thermorefractive process
\begin{equation}
\beta_{e,o}\approx-\frac{1}{n_{e,o}} \frac{dn_{e,o}}{dT}-\frac{1}{R} \frac{dR}{dT},
\end{equation}
where $R$ is the radius of the resonator. If the resonator is made out of z-cut stoichiometric lithium niobate, we find $(dn_{e}/dT)/n_e = 1.9\times10^{-5}$~K$^{-1}$ and $(dn_{o}/dT)/n_o = 7.1\times10^{-7}$~K$^{-1}$ at 25~$^o$C and 1064~nm optical wavelength \cite{jundt90jqe}. The thermal expansion coefficient is nearly identical for both optical modes and is determined by thermal expansion in the direction of an ordinary crystalline axis. It is $(dR/dT)/R = 1.5\times10^{-5}$~K$^{-1}$.
Similarly, for the acoustic wave we get
\begin{equation}
\beta_{s}\approx\frac{1}{V_s} \frac{dV_s}{dT}-\frac{1}{R} \frac{dR}{dT},
\end{equation}
where $V_s$ is the speed of sound in the material. Let us assume that we consider a surface acoustic wave propagating in the resonator. In this case $(dV_s/dT)/V_s \simeq -6\times10^{-5}$~K$^{-1}$ \cite{slobodnik71afcrl}.

The temperature change of the detuning parameter $\Delta_c$ is determined by the temperature shift of  optical mode frequencies $\Delta_c(\Delta T) \simeq (\beta_o-\beta_e) \Delta T \omega_a$. It is much larger than the acoustic shift. For instance, for lithium niobate we have $(\beta_o-\beta_e) \omega_a \simeq 5.2$~GHz/K ($\omega_a= 2.8 \times 10^{14}$~Hz for $\lambda_a=1064$~nm). On the other hand, $\beta_{s} \omega_c=7.5$~kHz/K if $\omega_c=0.1$~GHz.

To keep the opto-mechanical interaction phase matched one needs to keep
\begin{equation}
\Delta T \lesssim \frac{2(\gamma + \gamma_{ca})}{(\beta_o-\beta_e) \omega_a}=\frac{1}{Q_a(\beta_o-\beta_e)},
\end{equation}
where $Q_a$ is the optical quality factor. Assuming that $Q_a=10^9$ (the value achievable in stoichiometric crystals) we find that $\Delta T \lesssim 55\ \mu$K. Therefore, observing the power of the modulation sideband and applying  proper feedback to the resonator temperature to sustain the opto-mechanical oscillations one stabilizes the temperature of the resonator to submilliKelvin level. Going further and stabilizing the power level of the generated optical sideband by feeding it back into the resonator temperature, the resonator temperature can be stabilized to microKelvin level.

Another degree of freedom of the setup is detuning of the laser frequency from the eigenfrequency of the pumped mode. The temperature change of the detuning parameter $\Delta_a$ is given by quality of the laser lock to the optically pumped mode, as the frequency of the laser locked to the resonator mode follows the mode up to some extent. Since the temperature drift of the resonator is slow and the detuning is primarily determined by the internal drifts in the laser it is possible to keep $|\Delta_a|<0.1(\gamma + \gamma_{ca})$. A drift of $\Delta_a$ results in a shift of the maximum of the opto-mechanical gain, in accordance with Fig.~(\ref{fig1}). Estimating the maximum possible accuracy of stabilization of $\Delta_c$ we require the deviation of $\Delta_c$ to exceed the deviation of $\Delta_a$, $\delta|\Delta_c|>\delta |\Delta_a|$.  We find that $\delta|\Delta_c|\geq 0.1(\gamma + \gamma_{ca})$, if $|\Delta_a|<0.1(\gamma + \gamma_{ca})$. This condition allows achieving temperature stabilization better than several microKelvins for parameters specified above. Interestingly, the temperature stabilization also results in the increase of long term stability of the laser.

The thermal stabilization allows suppressing not only technical, but also fundamental thermodynamic fluctuations \cite{matsko07josab} of temperature within the mode volume in case of strong overlap between the optical and the acoustical modes. This is possible, for instance, in the case of light scattering on the surface acoustic waves \cite{matsko09prl}. The fundamental thermal fluctuations arise in the WGM volume even if the external temperature of the resonator is perfectly stabilized. In a small resonator with mode volume approaching $10^{-7}$~cm$^3$ the fundamental thermal fluctuations can reach microKelvin level. However, the fluctuations can be reduced by a proper low-temperature feedback, similarly to the feedback cooling achieved in opto-mechanics \cite{vinante08prl}.

To conclude, we have shown that opto-mechanical oscillators can be temperature stabilized using the strong temperature dependence of phase matching condition for the opto-mechanical process. The stabilization does not require presence of any external frequency and/or temperature reference. The method can be used for mechanical stabilization of the oscillator as the optical and mechanical modes respond differently to stress and strain. Finally, the method can be used for temperature stabilization of any nonlinear process (e.g. triply resonant frequency doubling \cite{ilchenko04prl}) characterized with a strong thermal dependence of the phase matching condition.



\end{document}